\documentclass[conference]{IEEEtran}
\IEEEoverridecommandlockouts
\usepackage{cite}
\usepackage{amsmath,amssymb,amsfonts}
\usepackage{algorithmic}
\usepackage{graphicx}
\usepackage{textcomp}
\usepackage{xcolor}
\usepackage{url}
\usepackage{amsmath,amssymb,amsfonts}
\usepackage{textcomp}
\usepackage{array}
\usepackage{color,framed}
\usepackage{multirow}
\usepackage{ulem}
\usepackage{algorithm}  
\usepackage{setspace}
\usepackage{comment}
\usepackage{pifont}
\usepackage{booktabs}
\newcommand{\tabincell}[2]{\begin{tabular}{@{}#1@{}}#2\end{tabular}}
\newcommand{\reffig}[1]{Fig.\ref{#1}}
\newcommand{\refeqs}[1]{Eq.\ref{#1}}

\def\BibTeX{{\rm B\kern-.05em{\sc i\kern-.025em b}\kern-.08em
    T\kern-.1667em\lower.7ex\hbox{E}\kern-.125emX}}
\begin{document}

\title{Dynamic Mining Interval to Improve Blockchain Throughput\\
}

\author{\IEEEauthorblockN{1\textsuperscript{st} Hou-Wan Long}
\IEEEauthorblockA{
\textit{The Chinese University of Hong Kong, Hong Kong}\\
1155190681@link.cuhk.edu.hk}
\and
\IEEEauthorblockN{2\textsuperscript{nd} Xiongfei Zhao}
\IEEEauthorblockA{
\textit{University of Macau, Macau}\\
yb97480@um.edu.mo}
\and
\IEEEauthorblockN{3\textsuperscript{rd} Yain-Whar Si}
\IEEEauthorblockA{
\textit{University of Macau, Macau}\\
fstasp@um.edu.mo}
}

\maketitle

\begin{abstract}

Decentralized Finance (DeFi), propelled by Blockchain technology, has revolutionized traditional financial systems, improving transparency, reducing costs, and fostering financial inclusion.  However, transaction activities in these systems fluctuate significantly and the throughput can be effected.  To address this issue, we propose a Dynamic Mining Interval (DMI) mechanism that adjusts mining intervals in response to block size and trading volume to enhance the transaction throughput of Blockchain platforms. Besides, in the context of public Blockchains such as Bitcoin, Ethereum, and Litecoin, a shift towards transaction fees dominance over coin-based rewards is projected in near future. As a result, the ecosystem continues to face threats from deviant mining activities such as Undercutting Attacks, Selfish Mining, and Pool Hopping, among others. In recent years, Dynamic Transaction Storage (DTS) strategies were proposed to allocate transactions dynamically based on fees thereby stabilizing block incentives.  However, DTS' utilization of Merkle tree leaf nodes can reduce system throughput. To alleviate this problem, in this paper, we propose an approach for combining DMI and DTS. Besides, we also discuss the DMI selection mechanism for adjusting mining intervals based on various factors.



\end{abstract}

\begin{IEEEkeywords}
    DeFi, block incentive, transaction fee, dynamic mining interval, mining target
\end{IEEEkeywords}

\section{Introduction}

Decentralized Finance, also known as DeFi, has seen a significant explosion of growth and development in recent years. This has been propelled by the power of Blockchain technology that has made it possible to recreate traditional financial systems in an open, decentralized manner, bypassing many of the limitations of conventional financial institutions. DeFi leverages Blockchain technology to offer services such as payment, asset exchange, lending and borrowing, and more, in a peer-to-peer, open-source manner. This has led to increased transparency, reduced costs, and the potential for financial inclusion, thereby contributing to its flourishing development. Transaction management is an important issue in both DeFi trading systems as well as in Cryptocurrency trading. In this paper, we consider these two different by overlapping issues.

In the context of Blockchain-based DeFi trading systems, such as payment systems, transaction volumes exhibit noticeable peaks and valleys in correlation with changes in trading hours. High trading volumes can typically be observed during meal periods, with a decline in activity at night. This suggests that transaction activities in these systems are heavily influenced by human behavior and daily routines. In traditional securities trading systems, a similar pattern of trading peaks and troughs can be observed at key time nodes, such as market opening and closing times. However, these fluctuations can adversely effect the throughput of the Blockchains.

\begin{figure}[H]
    \makeatletter
    \def\@captype{figure}
    \makeatother
    \centering
    \includegraphics[width=0.42 \textwidth]{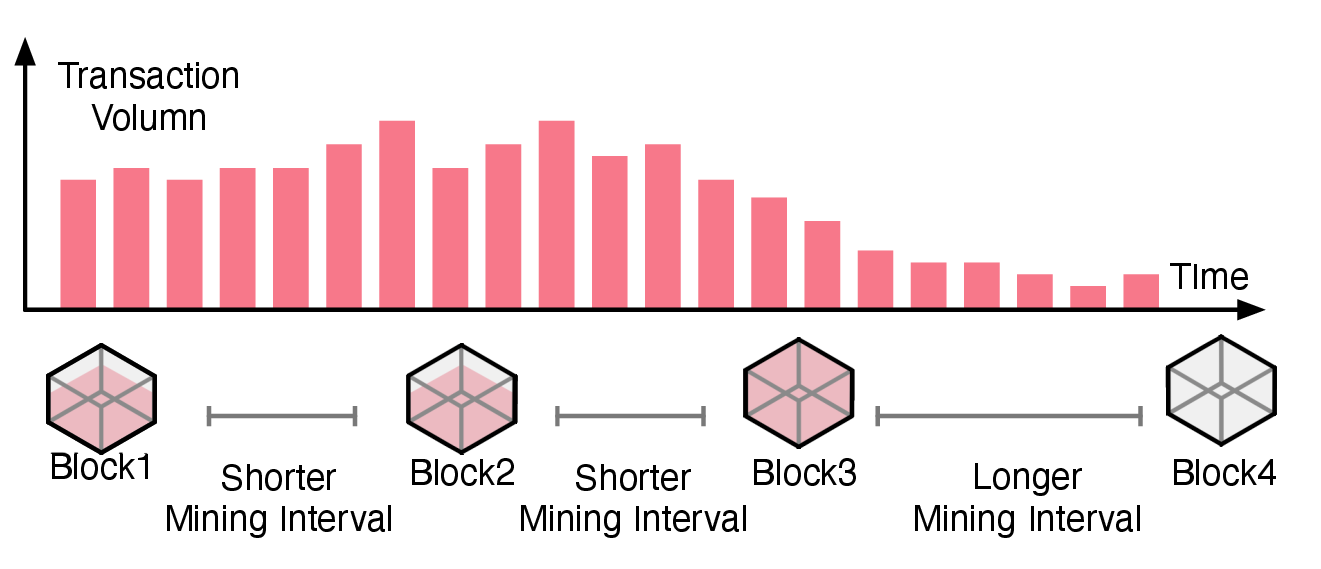}
    \caption{Dynamic Mining Interval (DMI) mechanism adjusts the mining interval in response to transaction volumes to address the fluctuating transaction volumes in Blockchain-based DeFi trading systems.}
    \label{TransactionsDMI}
\end{figure}

To address the fluctuation in transaction volumes of Blockchain-based DeFi trading systems, we propose a novel mechanism called Dynamic Mining Interval (DMI). \textcolor{black}{In this paper, mining interval is defined as the expected time to mine a new block since the last block is mined. For instance, the mining interval of current Bitcoin network is fixed at 600s.} DMI adjusts the mining interval in response to the blocksize. The miner who successfully solves the computational mining problem updates the target value in the current block, which is contingent upon the block size, thereby facilitating a responsive adaptation to transaction activity levels. As illustrated in \reffig{TransactionsDMI}, during periods of off-peak trading volume, the mining interval is shortened by DMI. This allows for the generation of more blocks and improves the overall transaction throughput of the Blockchain platform, ensuring efficient processing of transactions. Conversely, during periods of low transaction volume, the mining interval is lengthened by DMI. This helps reduce the fork rate of the Blockchain, keeping the network safe. Specifically, the proposed DMI mechanism aims to effectively manage the fluctuating transaction volumes in Blockchain-based Decentralized Finance (DeFi) systems, enhancing both their efficiency and scalability. Besides, the scope of the DMI mechanism extends further. DMI can be employed to augment the processing throughput in Blockchain-based transaction processing systems. In these systems, transaction fees are mainly used as mining rewards.

\vspace{-0.4cm}


In the context of Cryptocurrency such as Bitcoin network, the coin-based reward is still the current primary revenue source, even though it is preset to be halved approximately every four years. The expectation is that transaction fees will gradually replace coin-based rewards as the primary revenue source \cite{EASLEY201991}. However, deviant mining threats have emerged under the transaction fee regime,  such as the Undercutting Attack \cite{carlsten2016}, Selfish Mining \cite{eyal2013majority}, Pool Hopping \cite{rosenfeld2011analysis} and Mining Gap \cite{carlsten2016} which can significantly impact the system's integrity and efficiency. To address these challenges, Zhao et al. \cite{zhao2022dynamic} have proposed the concept of Dynamic Transaction Storage (DTS) strategies. Despite the significant progress made by the DTS strategy in addressing the challenges of transaction-fee regimes in Blockchain systems, there are still potential issues that must be considered. One key issue is the utilization of Merkle tree leaf nodes in the DTS strategy, which can result in empty nodes within the tree. This can adversely affect the overall throughput of the system, leading to a reduction in the number of transactions that can be contained in a block. 

\begin{figure}[H]
    \makeatletter
    \def\@captype{figure}
    \makeatother
    \centering
    \includegraphics[width=0.42 \textwidth]{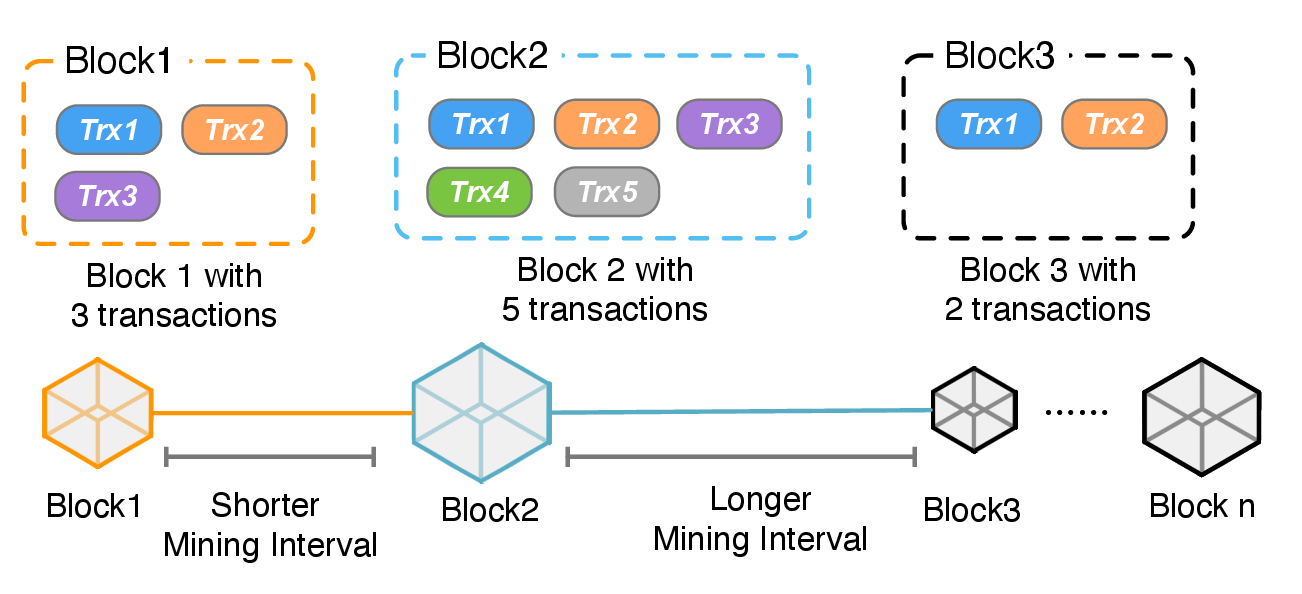}
    \caption{DTS strategy combined with DMI mechanism to dynamically regulate the mining intervals based on variations in block size.}
    \label{DMI4DTS}
\end{figure}

To address the potential drawbacks of the DTS strategy and mitigate the negative impacts on system performance, in this paper, we propose an approach to combine DTS strategy with DMI mechanism. DTS strategy in a Blockchain can result in generation of blocks with varying sizes containing different number of transactions. This presents an opportunity to leverage the DMI mechanism to dynamically regulate the mining interval based on the block size. \textcolor{black}{As illustrated in \reffig{DMI4DTS}, a miner engaged in mining a smaller size block, such as Block 1 or Block 3, sets a higher target value in comparison to a larger size Block 1. This approach recognizes the differing complexities and workloads associated with blocks of varying sizes. Operating on the principle that all miners work on the longest chain, they collectively share this current target value which facilitates a reduced mining interval for Blocks 1 and 3, while extending it for Block 2.} The miner who successfully solves the computational mining problem updates the target value in the current block, which is contingent upon the current block size. By adjusting the mining interval in response to changes in block size, the DMI mechanism can optimize the block generation process and enhance the efficiency of the Blockchain system.

\vspace{-0.4cm}

\vspace{-0.3cm}
To this end, this paper provides a detailed assessment of the performance improvement achieved by the DMI mechanism in addressing the payment transaction volume fluctuations for the Industrial and Commercial Bank of China (ICBC). Furthermore, we also evaluate the effectiveness of DMI on DTS strategy. Experimental results show that DMI with DTS strategy can achieve significant enhancements in transaction throughput while maintaining a stable fork rate. Our contributions can be summarized in three aspects: 


\begin{itemize}
    \item[a)] In this paper, we propose a novel Dynamic Mining Mechanism (DMI) to enhance the throughput of Blockchains.
    
    \item[b)] DMI is also integrated with DTS strategy to further enhance the transaction throughput while maintaining a stable fork rate.
    
    \item[c)] We utilize payment transaction data from the Industrial and Commercial Bank of China (ICBC) to evaluate the effectiveness of DMI mechanism. Our results show that DMI can significantly improve the transaction processing capacity of Blockchain platforms during busy hours while reducing network and computing resource waste during transaction troughs.

    \item[d)] We also verify that DMI can mitigate the negative impact of DTS strategy on throughput. Specifically, by integrating DMI and DTS, we are able to maintain stable block incentives and achieves greater throughput in scenarios where block rewards are based solely on transaction fees.

    \item[e)] We propose the selection mechanism of DMI for adjusting the mining interval based on different factors. The mechanism allows whether to increase the throughput by dynamically adjusting the computing power of the entire Blockchain network or to save resources.
\end{itemize}

This paper is organized as follows. Section 2 summarizes the related work. Section 3 introduces the preliminaries. In Section 4, we describe the framework for DMI mechanism. Section 5 illustrates the experiment settings and experimental outcomes. The paper is concluded in Section 6.

\section{Related work}

Our research on the DMI mechanism is related to several prior studies. The first is DTS strategy, which aims to stabilize block incentives and prevent deviant mining behavior. However, DTS strategy reveals a trade-off between block incentive volatility and Blockchain throughput. The second set of studies explores the difficulty adjustment algorithm (DAA). Our research builds on these works by proposing an algorithm to control the mining interval under DMI based on the total network hash rate and block size.

\subsection{Dynamic Transaction Storage (DTS) strategy}

In \cite{zhao2022dynamic}, Zhao et al. proposed the concept of Dynamic Transaction Storage (DTS) strategy to address the issue of deviant mining strategies that can occur when switching to a transaction-fee regime. These deviant mining strategies include Undercutting Attack, Selfish Mining, and Pool Hopping \cite{carlsten2016instability}. Under the transaction-fee regime, block incentives rely solely on transaction fees, which are not fixed and can fluctuate depending on the level of network congestion and market activity. These fees are influenced not only by the amount of currency contained in the transaction but also by the level of market activity during different time periods. In cases where there is a high volume of transactions within a short period of time, the willingness of users to pay higher fees encourages miners to prioritize and process their transactions more quickly.

\begin{figure}[H]
    \makeatletter
    \def\@captype{figure}
    \makeatother
    \centering
    \includegraphics[width=0.42 \textwidth]{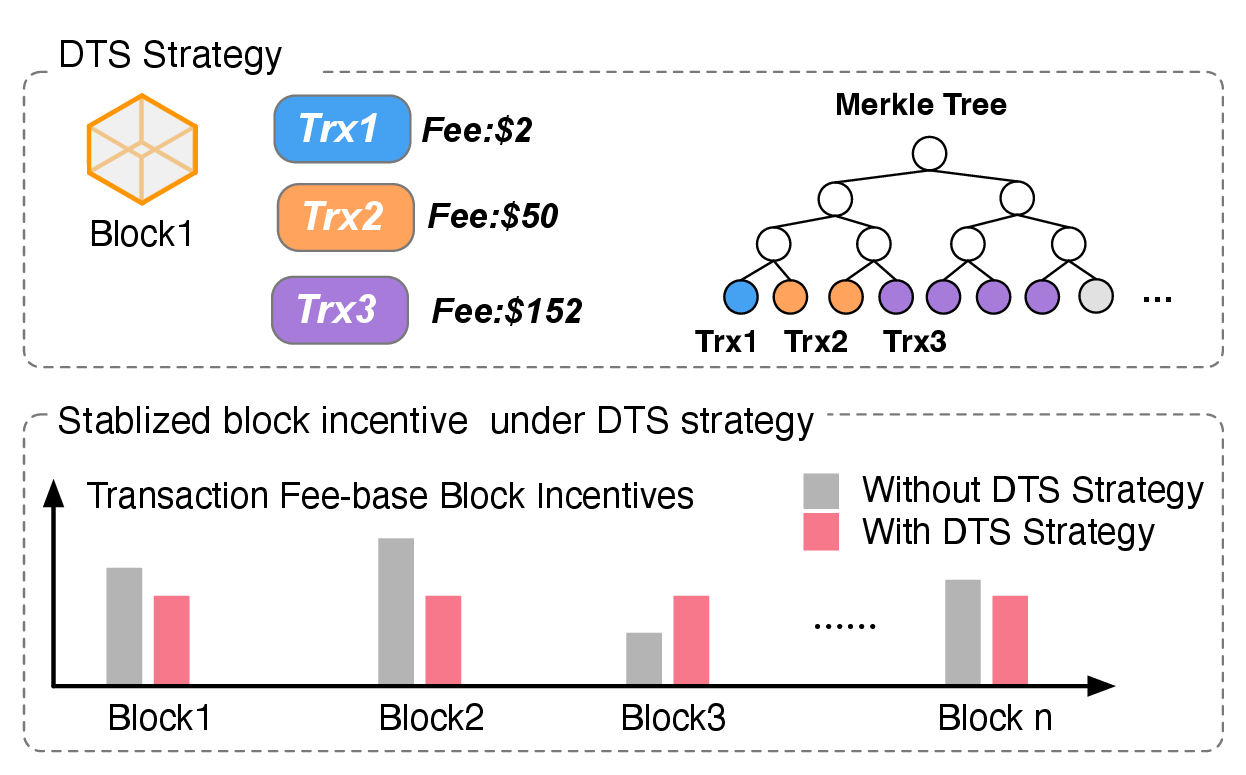}
    \caption{DTS strategy \cite{zhao2022dynamic} allocates transactions in a Merkle Tree data structure based on transaction fees to ensure stable block incentives.}
    \label{DTS}
\end{figure}

\textcolor{black}{As depicted in the ``DTS Strategy'' section of \reffig{DTS}, miners adopt the DTS strategy for each transaction selected from the mempool. This strategy employs the Cumulative Distribution Function (CDF) to calculate the number of Merkle Tree leaf nodes each transaction occupies based on its fee level. Transactions with higher fees are allocated more leaf nodes within a block, while transactions with lower fees are assigned fewer nodes. Upon the allocation of all Merkle Tree leaf nodes, miners generate the block and propagate it to the rest of the Blockchain members. In the ``Stabilized Block Incentive Under DTS Strategy'' section of \reffig{DTS}, the key objective of the DTS strategy is to convert the volatile nature of transaction fees into a predictable and stable source of block incentives. This transformation contributes to a more reliable and sustainable block incentive within the Blockchain.}


\vspace{-0.4cm}

This strategy offers a viable solution to the issue of strategic deviations which are likely to occur during the transition to a transaction-fee-dominated regime. Through dynamic allocation of transactions based on their respective fees, DTS ensures that miners are incentivized to include the transactions while maintaining the integrity and efficiency of the network. However, according to the experiment result conducted in our study, pursuing low volatility in block incentives through DTS is at the expense of resulting in drastic oscillation in block size, which has a negative effect on Blockchain throughput \cite{Singh2020MultiobjectiveOO}. 

\subsection{Blockchain difficulty adjustment algorithm (DAA)}
Adjusting the block generation interval dynamically on the basis of utilizing fluctuating block size is the core of this study. However, arbitrarily shortening the block generation interval can increase the fork rate \cite{fork_rate_analysis2019} and jeopardize Bitcoin sustainability. Therefore, determining the relation between block size, fork rate and difficulty control is an indispensable part in the process of finding an appropriate algorithm that generate adaptive block mining interval with respective to block size. There exist many works in the literature which address difficulty adjustment algorithm for Bitcoin.

\begin{itemize}
    \item Kraft et al. \cite{Kraft2016DifficultyCF} proposed a difficulty update mechanism, which is based on modeling the mining process using Poisson processes and designing a new difficulty update algorithm. However, the algorithm proposed in \cite{Kraft2016DifficultyCF} is elaborated to address the instability of hash rate, while DMI mechanism proposed in this paper \textcolor{black}{dynamically adjusts its target in accordance with the latest block size}.
    \item Feng et al. \cite{RTPoW} introduced RTPoW, a consensus protocol featuring a real-time difficulty adjustment algorithm. This mechanism facilitates the adjustment of each block's difficulty target by predicting real-time total hash rate, thereby ensuring block time stability even amidst substantial hash rate fluctuations. However, in contrast to our approach, RTPoW's modification on the target, which is solely based on hashrate predictions, does not address the instability originating from fluctuating block sizes. 
    
    \item \textcolor{black}{Cao et al. \cite{metaRegulation} introduced Meta-Regulation, a method that enhances throughput by adjusting mining difficulty after each time window of $n$ blocks, while maintaining a fixed fork rate of $2\%$. However, Meta-Regulation lacks the ability to promptly respond to spikes in DeFi transactions and the volatile block size of DTS, as its adjustments are solely based on time windows. In contrast, the DMI mechanism we propose effectively addresses these challenges by adjusting mining difficulty after a new block is mined. Furthermore, while the fork rate baseline of $2\%$ in Meta-Regulation is arbitrarily set, the DMI mechanism derives its fork rate baseline through rigorous experimentation.}
    
    \item Kanda et al. \cite{BlockIntervalAdjustment} introduced a Block Interval Adjustment (BIA) algorithm designed to regulate the mining interval. This algorithm modifies the block interval contingent on the fork rate. However, BIA does not optimize throughput when the fork rate is low. Conversely, the DMI mechanism takes into account block size fluctuations to capitalize on periods of low fork rate, thereby enhancing throughput.
\end{itemize}

Next, we will describe the specific working mechanism of DMI, which aims to adjust the mining interval based on blocksize, fork rate and hash rate.

\section{Dynamic Mining Interval (DMI) Mechanism}

The DMI algorithm aims to dynamically control the expected mining interval to improve throughput while ensuring that the estimated fork rate remains within a predetermined acceptable threshold. Note that Blockchain networks rely on the target function to regulate the difficulty of mining new blocks and ensure that blocks have sufficient time to propagate through the network, thereby maintaining the stability and security of the system. The proposed DMI mechanism improves upon the current target function by adjusting the target value for every block based on the size of the newly mined block. This modification enhances the current method of adjusting the difficulty of every 2016 block. The modified target function sets a reasonable target value for the next miner to mine the interval just long or short enough for the current block to propagate throughout the Blockchain network. By adopting this approach, the DMI mechanism can dynamically control each expected mining interval to improve the throughput while simultaneously maintaining the estimated fork rate within an acceptable threshold. For the purpose of clarification, all notations used in this paper are listed in Table.~\ref{notations}

\vspace{-0.3cm}
\linespread{1}
\begin{table}[H]
    \centering
    \footnotesize
    \caption{\textcolor{black}{Notations used in this paper}}
    \resizebox{0.96\hsize}{!}{
        \begin{tabular}{p{1cm}<{\raggedright} | p{6.5cm}<{\raggedright}}
            \hline\hline
            \specialrule{0em}{3pt}{1pt}
            \textcolor{black}{\textbf{Notation}} & \textcolor{black}{\textbf{Definition}} \\
            \specialrule{0em}{3pt}{1pt}
            \hline  
            \specialrule{0em}{1pt}{1pt} 
            \textcolor{black}{$N$} & \textcolor{black}{node in the Blockchain network} \\
            \specialrule{0em}{1pt}{1pt}
            \hline  
            \specialrule{0em}{1pt}{1pt} 
            \textcolor{black}{$TP$} & \textcolor{black}{block propagation time between designated 2 nodes} \\
            \specialrule{0em}{1pt}{1pt}
            \hline  
            \specialrule{0em}{1pt}{1pt} 
            \textcolor{black}{$B$}  & \textcolor{black}{newly mined block}\\
            \specialrule{0em}{1pt}{1pt}
            \hline  
            \specialrule{0em}{1pt}{1pt} 
            \textcolor{black}{$t$}  & \textcolor{black}{time after new block is mined, among $[0, 10]$} minutes\\
            \specialrule{0em}{1pt}{1pt}
            \hline  
            \specialrule{0em}{1pt}{1pt} 
            \textcolor{black}{$n$} & \textcolor{black}{number of nodes receive the block within time $t$} \\
            \specialrule{0em}{1pt}{1pt}
            \hline  
            \specialrule{0em}{1pt}{1pt} 
            \textcolor{black}{$m$} & \textcolor{black}{number of neighbour nodes} \\
            \specialrule{0em}{1pt}{1pt}
            \hline  
            \specialrule{0em}{1pt}{1pt} 
            \textcolor{black}{$P$} & \textcolor{black}{probability of generating new block} \\
            \specialrule{0em}{1pt}{1pt}
            \hline
            \specialrule{0em}{1pt}{1pt} 
            \textcolor{black}{$I$} & \textcolor{black}{mining interval} \\
            \specialrule{0em}{1pt}{1pt}
            \hline
            \specialrule{0em}{1pt}{1pt} 
            \textcolor{black}{$f(t)$} & \textcolor{black}{informed nodes rate function} \\
            \specialrule{0em}{1pt}{1pt}
            \hline
            \specialrule{0em}{1pt}{1pt} 
            \textcolor{black}{$r$} & \textcolor{black}{fork rate} \\
            \specialrule{0em}{1pt}{1pt}
            \hline
            \specialrule{0em}{1pt}{1pt} 
            \textcolor{black}{$r_0$} & \textcolor{black}{baseline fork rate for DMI} \\
            \specialrule{0em}{1pt}{1pt}
            \hline
            \specialrule{0em}{1pt}{1pt} 
            \textcolor{black}{$D$} & \textcolor{black}{mining difficulty} \\
            \specialrule{0em}{1pt}{1pt}
            \hline
            \specialrule{0em}{1pt}{1pt} 
            \textcolor{black}{$R$} & \textcolor{black}{hash rate} \\
            \specialrule{0em}{1pt}{1pt}
            \hline
        \end{tabular}
    }
    \label{notations}
\end{table}

The DMI mechanism comprises several steps:     

\textbf{Step 1. Fork rate limit specification}: 
Owing to the distributed nature of the network and communication speed limitations, delays may arise between the moment a block is mined and the time it is received by all nodes in the network. These delays can result in the creation of forks in the Blockchain, wherein different nodes possess alternate versions of the Blockchain due to receiving distinct blocks at varying times. The rate at which forks arise owing to propagation delays is referred to as ``fork rate''. A high fork rate can compromise the security and reliability of the Blockchain, as it may either hinder or prevent nodes from reaching a consensus on the Blockchain's true state. Consequently, minimizing the fork rate is a critical objective for Blockchain networks. Decker et al. \cite{propagationdecker} posited that the fork rate, \textit{r}, can be estimated using the following equation:

\begin{equation}
    r=1-(1-P)^{\int_{0}^{\infty}{(1-f(t))}\mathrm{d}t} \qquad P=\frac{1}{I} \label{frokRate}
\end{equation}

\begin{figure}[htp]
    \makeatletter
    \def\@captype{figure}
    \makeatother
    \centering
    \includegraphics[width=0.4 \textwidth]{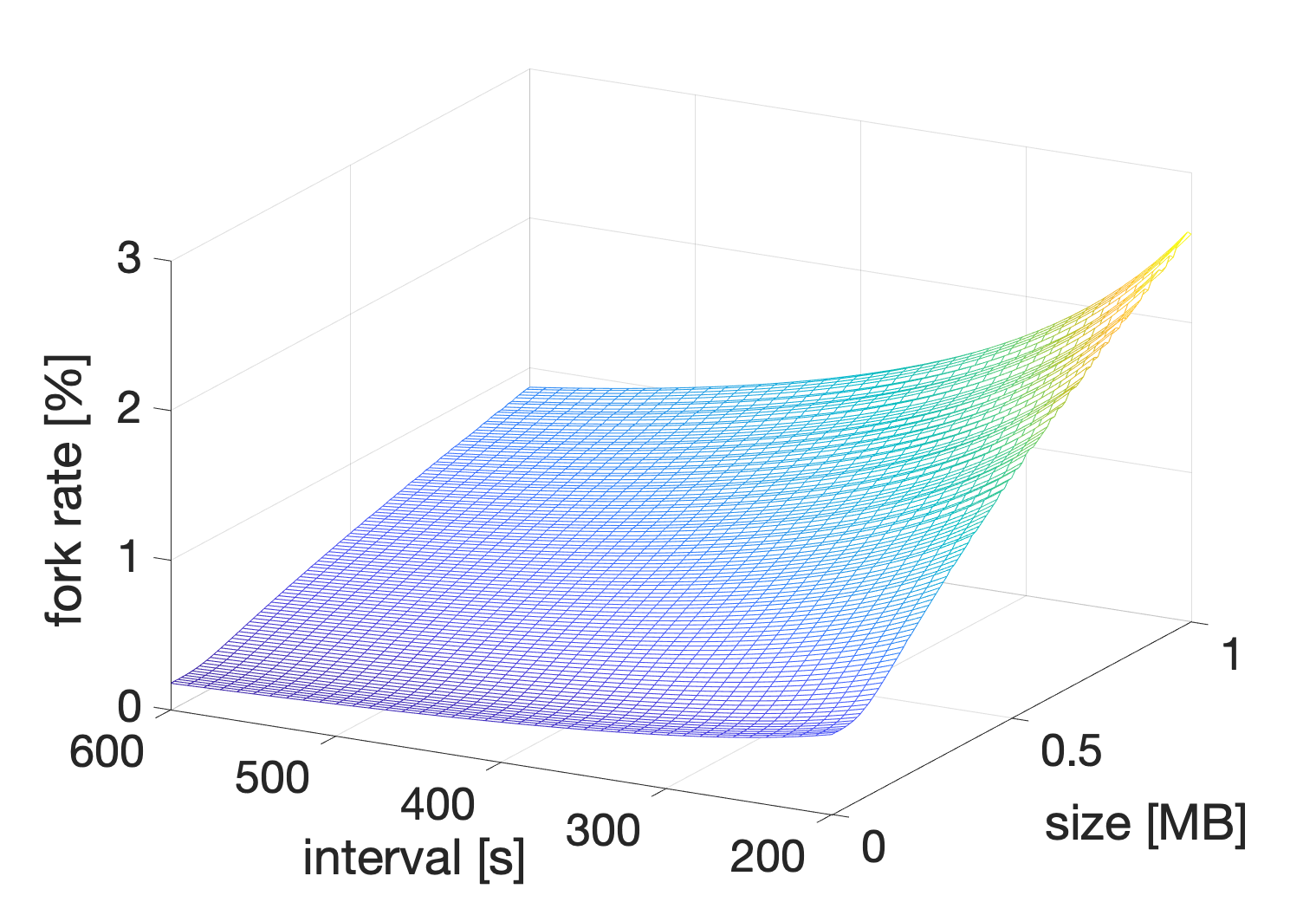}
    \caption{Three-dimensional diagram of interval, block size and fork rate, graphical illustration of Eq.1}
    \label{Block size vs Interval vs Fork rate}
\end{figure}

In this equation, \textit{P} represents the probability of the network as a whole discovering a block in any given second, while \textit{f(t)} denotes the informed nodes rate function, as defined in Section 3.4.
Prior to the mining process, DMI sets a fork rate limit, denoted by \textit{$r_0$}. This parameter is crucial as it defines the acceptable level of risk associated with temporary forks in the Blockchain. By setting an appropriate value for \textit{$r_0$}, the DMI mechanism can effectively manage the trade-off between mining efficiency and the potential for network disruption due to excessive forking.
By using \refeqs{frokRate}, we can determine the relationship amongst block size, mining interval, and fork rate (see Fig.\ref{Block size vs Interval vs Fork rate}).

\textbf{Step 2. Informed nodes rate function generation}: During the mining process, the miner receives an informed note rate function (see \refeqs{informedNodes}) that represents the rate of informed nodes in the network by taking into account parameters such as network latency and block size. 
In a decentralized Blockchain network such as Bitcoin, nodes rely on the dissemination of information to attain consensus on the state of the Blockchain. Upon mining a new block, a node broadcasts the block to its peers within the network. These peers subsequently validate and propagate the block to their own peers. This process iterates until all nodes in the network receive the block and reach a consensus on its validity, allowing its addition to the Blockchain. The informed nodes rate, denoted as \textit{f(t)}, can be calculated based on the following formula:

\begin{equation}
    f(t)=\frac{n}{N} \label{informedNodes}
\end{equation}

where \textit{n} denotes the number of nodes aware of a newly mined block at time \textit{t}, while \textit{N} denotes the total number of nodes in the network. By definition, informed nodes rate is a function of time \textit{t} which suffices the following properties:
\begin{enumerate}
    \item[a)] {increasing on domain $[0,\infty)$}, and
        \item[b)] {bounded below by 0 and bounded above by 1.}
\end{enumerate}

\begin{figure}[htp]
    \makeatletter
    \def\@captype{figure}
    \makeatother
    \centering
    \includegraphics[width=0.36 \textwidth]{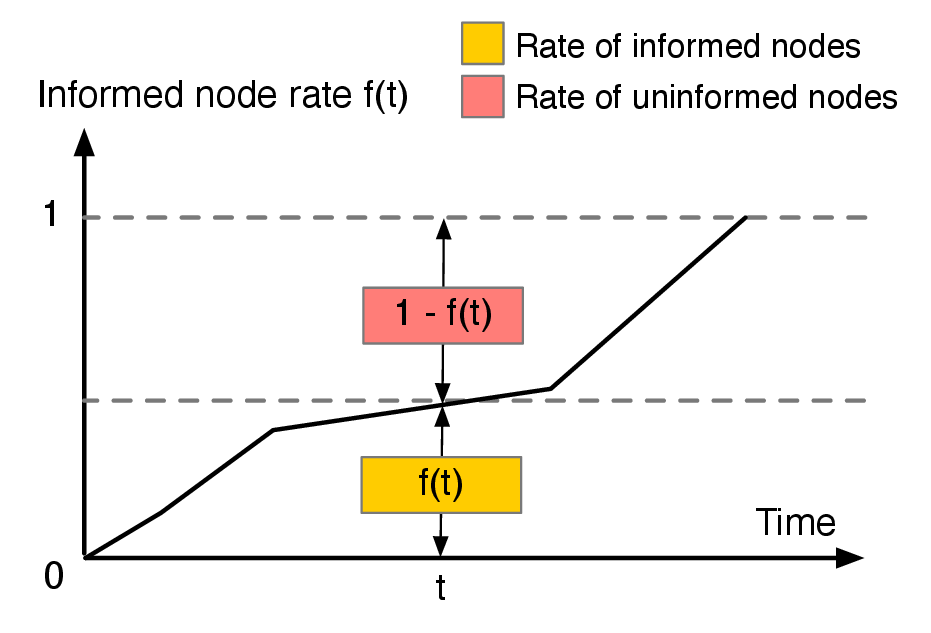}
    \caption{Illustration of informed nodes rate}
    \label{Informed nodes rate}
\end{figure}

These properties are illustrated in Fig.\ref{Informed nodes rate}.

For simplicity, we provides a probabilistic model of how new blocks spread in a Blockchain network with a given node degree, network size, blocksize, bandwidth, and delay to determine \textit{f(t)}. The network consists of \textit{N} nodes, with each node connected to \textit{m} neighbor nodes. When a new block is discovered by a node, it is propagated across the network through these neighbor connections in a cascading manner. To model this propagation, we make the simplifying assumption that nodes randomly select $m$ neighbors to propagate new blocks to, with replacement. This means that there is a possibility that a node may receive the block multiple times, or that not all of its $m$ neighbors receive the block in a given propagation wave. However, over time, the block will eventually reach all nodes in the network.

\textcolor{black}{We use the term \textit{$P_k$} to represent the probability that a neighboring node has not yet received the new block at the \textit{kth} wave. Initially, \textit{$P$} is set to 1 since only one node possesses the new block (i.e., \textit{$P_1=1$}). In the first wave of propagation, the node that discovered the block shares it with its \textit{m} neighboring nodes. This results in \textit{m+1} nodes possessing the block, while \textit{N-m-1} nodes still do not have it.}

\textcolor{black}{The time interval between two consecutive propagation waves is denoted as \textit{$TP_{avg}$}. To account for realistic network delays, we define \textit{$TP_{avg}$} as ($\frac{Blocksize}{Bandwidth}$ + Delay), where Blocksize represents the size of the propagated block, Bandwidth denotes the average network bandwidth, and Delay signifies the average network delay.}

\textcolor{black}{During the second wave, each of the \textit{m} nodes that received the block in the first wave propagates it to \textit{m} neighboring nodes. The probability that a neighbor needs to receive the block is denoted as \textit{$P_2$} and $P_2=\frac{N-mP_1}{N}$. In this wave, we can expect approximately \textit{$m^2P_2$} new nodes to receive the block. Consequently, after this wave, there will be \textit{$1+mP_1+m^2P_2$} nodes possessing the block, while \textit{$N-1-mP_1-m^2P_2$} nodes remain without it.}

\textcolor{black}{This process continues, with the probability \textit{$P_k$} decreasing by a factor of \textit{$\frac{N-nodes_informed}{N}$} after each wave. The repetition continues until the probability \textit{$P$} reaches 0, signifying that all N nodes in the network have received the new block.}

\textbf{Step 3. Calculation of expected interval}: The expected interval between block arrivals is calculated using \refeqs{frokRate}. This calculation incorporates the informed nodes rate function generated in the previous step and is a key factor in determining the appropriate mining difficulty for the upcoming block. By adjusting the expected interval based on real-time network conditions, the DMI mechanism ensures that the mining process remains both efficient and secure.

\textbf{Step 4. Mining difficulty level conversion}: Kraft et. al \cite{Kraft2016DifficultyCF} demonstrated that under the circumstance of a constant hash rate, the mining process is a homogeneous Poisson process, and the expected mining interval ($I$) can be calculated based on the following equation:

\vspace{-0.3cm}
\begin{equation}
    I=\frac{D}{R} \label{MiningInterval}
\end{equation}

\noindent where hash rate \textit{R} represents the number of hashes the network can execute per second, and difficulty \textit{D} represents the expected number of hashing calculations needed for a newly mined block.
The miner then converts the expected interval into the appropriate mining difficulty level using Equation \refeqs{MiningInterval}. This step is essential for maintaining a consistent level of mining competition throughout the network, which in turn helps to prevent any single miner or group of miners from gaining a disproportionate amount of influence over the Blockchain.

\textbf{Step 5. Calculation of target value}: 
Generally, the target is commonly referred to as the difficulty level, as it provides an indication of the complexity involved in the mining process. For an arbitrarily chosen nonce, the probability that the hash value is smaller than \textit{T} is $\frac{T}{2^{256}}$. Then *D* can be calculated based on the following equation:
\begin{equation}
    D=\frac{2^{256}}{T}\label{MiningDiff2}
\end{equation}
Conversely, \textit{D} can be used to deduce \textit{T},
\begin{equation}
    T=\frac{2^{256}}{D} \label{MiningDiff}
\end{equation}

\textcolor{black}{By combining Eq.1, Eq.3, and Eq.5, we derive the following equation. This equation serves as a guideline for miners to determine the appropriate target value for the next block in accordance with block size $s$ and baseline fork rate $r_0$.}

\begin{equation}
    T(r_0,s)=\frac{2^{256}(1-{(1-r_0)}^\frac{1}{\int_0^\infty{(1-f(t))}\mathrm{d}t})}{R}
\end{equation}

\vspace{-0.3cm}
\begin{algorithm}
\small
\caption{Calculation of Target \textit{T} in DMI Mechanism}
    \textbf{Input:} Fork rate limit $r$, Block size $S$
    \textbf{Output:} Target $T$
    \textbf{Step 1:} Specify forking rate limit $r$
    \textbf{Step 2:} Calculate informed nodes rate $f(t)$ using \refeqs{informedNodes}
    \begin{align*}
        f(t) = \frac{\text{Number of informed nodes at t}}{\text{Number of nodes}}
    \end{align*}
    \textbf{Step 3:} Calculate expected interval $I$ using \refeqs{MiningInterval} and \refeqs{frokRate}
    \begin{align*}
        I = \frac{1}{1-{(1-r)}^{\frac{1}{\int^\infty_0(1-f(t))\mathrm{d}x}}}
    \end{align*}
    \textbf{Step 4:} Convert expected interval $I$ to mining difficulty $D$ using \refeqs{MiningDiff2}
    \begin{align*}
        D = I.R
    \end{align*}
    \textbf{Step 5:} Calculate target $T$ using \refeqs{MiningDiff}
    \begin{align*}
        T = \frac{{2^{256}}}{D}
    \end{align*}
    \textbf{return} Target $T$
\end{algorithm}

\textbf{Step 6. Target insertion}

\begin{figure}
\small
    \centering
    \includegraphics[width=0.42 \textwidth]{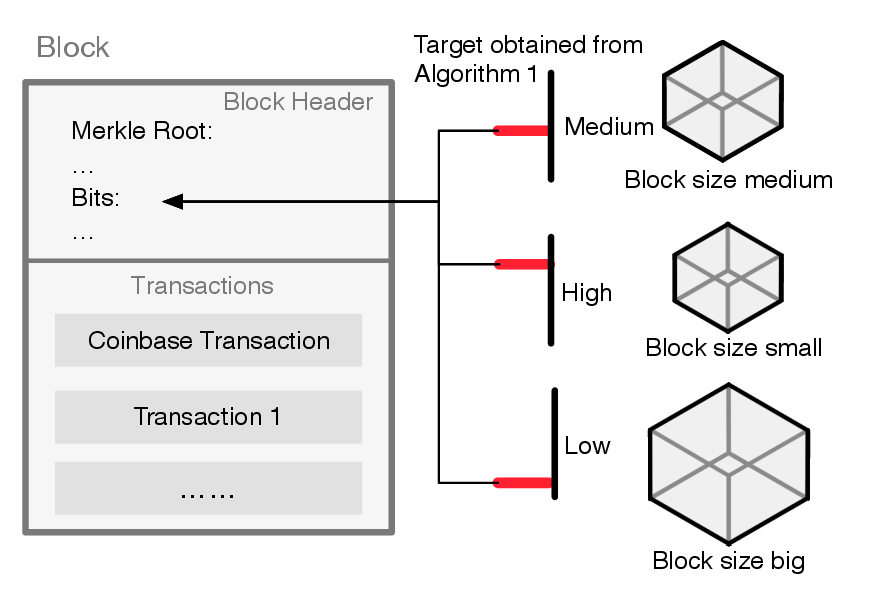}
    \caption{Blocks are written with different targets with respect to their size}
    \label{DMI_3}
\end{figure}

\textcolor{black}{The target value obtained from Algorithm 1 is stored in the block header's ``Bits''. This integration effectively lays out the mining challenge that must be met by other miners to successfully mine the subsequent block. As illustrated in \reffig{DMI_3}, the miner assigns target values with a strategic approach: a target value of medium difficulty for a medium-sized block, a higher target value for a smaller block, and a lower target value for a larger block. This strategy aids in controlling the mining interval of the next block, with the goal being to provide sufficient time for the current block to propagate throughout the entire network. This adjustment ensures an efficient and balanced overall mining process, thereby contributing to the robustness and performance of the blockchain network.}

\textcolor{black}{To substantiate the efficacy of the Dynamic Mining Interval (DMI) mechanism, we will present a series of simulation experiments conducted across diverse scenarios in the next section.}

\section{Experiments}

We have conducted three comprehensive sets of simulation experiments to evaluate the effectiveness of the proposed DMI mechanism. The first set of experiments is designed to evaluate DMI's impact on DeFi and to investigate potential solutions to solve the congestion during payment's peak hours. The second set of experiments is designed to evaluate DMI's impact on DTS, aiming at eliminating the downside of DTS strategies and improving its throughput. The third set of experiments is used to compare DMI with other mechanisms. 

\vspace{-0.3cm}
\subsection{Experiment settings}

To ensure that our experiments are aligned with the current status of Bitcoin, we have listed the parameter settings in Table \ref{SIMULATION SETTINGS BASED ON THE BLOCKCHAIN}. The experiment for the impact of DMI on DeFi is based on 347071 payment data from July 2023 to August 2023, which is obtained from the Industrial and Commercial Bank of China (Macau) Limited (ICBC Macau). The experiment for the impact of DMI on DTS is based on historical market data from January 2017 to December 2020, which includes 2,000,000 transactions with minute-to-minute updates of OHLC (Open, High, Low, Close), transaction amount in BTC, and weighted Bitcoin price. To establish the baseline forking rate applied in DMI, we conducted a simulation using the default network setting in SIMBLOCK, consisting of 10,000 blocks with a size of 1MB. 

\begin{table}[htp]
\small
    \centering
    \caption{SETTINGS FOR SIMULATION EXPERIMENT}
    \begin{tabular}{ c|c }
    \hline\hline
    \specialrule{0.00em}{3pt}{1pt} 
    \textbf{Parameter}     & \textbf{Value} \\
    \specialrule{0.00em}{3pt}{1pt} 
    \hline
    \specialrule{0.00em}{1pt}{1pt} 
    Transaction     & 500 Byte \\
    \specialrule{0.00em}{1pt}{1pt} 
    \hline
    \specialrule{0.00em}{1pt}{1pt} 
    Block Size & 1 MB \\
    \specialrule{0.00em}{1pt}{1pt} 
    \hline
    \specialrule{0.00em}{1pt}{1pt} 
    Nodes Number & 10000 \\
    \specialrule{0.00em}{1pt}{1pt} 
    \hline
    \specialrule{0.00em}{1pt}{1pt} 
    Transactions in Mempool & 16,000 Transactions \\
    \specialrule{0.00em}{1pt}{1pt} 
    \hline
    \specialrule{0.00em}{1pt}{1pt} 
    Hash Rate & 40000000\\
    \specialrule{0.00em}{1pt}{1pt} 
    \hline
    \toprule
    \end{tabular}
    \label{SIMULATION SETTINGS BASED ON THE BLOCKCHAIN}
\end{table}

By utilizing Bitcoin's historical market data and adopting parameter settings that align with the current Bitcoin status, we aim to ensure that our experiments are relevant and reliable. These efforts contribute to the validity and applicability of our findings. We set five key parameters along with their respective values for our simulation. The transaction size is set at 500 bytes, which is the data unit size the simulator will use for each transaction. The hash rate, designated as $R$, is set to 4,000,000 hashes per second, indicating the speed at which computations can be performed. The number of nodes, represented as $N$, is set to 10,000, demonstrating the total number of active nodes in the simulation. Each node is connected to 8 neighbor nodes, as indicated by $m$. Lastly, the baseline forking rate, denoted as $r_0$, is set at 0.95\%, which is the basic rate at which new branches are created in the simulation.

\subsection{Benchmarks}

In this study, we utilize throughput as a benchmark for evaluating the scalability of the processing cability improvement of Blockchain network under DMI mechanism. 

\begin{itemize}
    
    \item \textbf{Throughput} is a performance metric that measures the effectiveness of transactions confirmed per second (TPS) by the Blockchain. In our experiments, we measure the throughput to evaluate the effectiveness of the Blockchain network as a viable payment alternative. With Bitcoin generating a 1 MB block in a 10-minute interval, users typically have to wait for six blocks to obtain the final state of a transaction. This results in a low throughput of approximately 3.5 transactions per second, which is far less than the throughput of VISA or Paypal. Therefore, a higher throughput is essential for wider adoption of Blockchain technology as a payment alternative.
    
    
    \item \textbf{Forking rate} is another important metric for our experiments, as it affects the stability of the Blockchain network. Forking occurs when multiple miners find valid blocks simultaneously, resulting in the creation of multiple versions of the Blockchain. High forking rates can lead to network instability and undermine the integrity of the Blockchain network. Therefore, forking rate must be carefully considered when evaluating the performance of the Blockchain network. In our experiments, we measure the forking rate to evaluate the effectiveness of the proposed mechanism in maintaining network stability while improving mining efficiency.
    
\end{itemize}

\subsection{Experiment 1: Applying DMI for DeFi Payment}
The utilization of DMI is aimed at increasing the generation of blocks when they are not fully filled, potentially optimizing the Blockchain's performance during high-demand periods. Subsequent to the DMI adjustment, we simulate the processing of payment transactions using a tool named 'SIMBLOCK.' SIMBLOCK is employed to emulate the transaction processing environment and measure the overall throughput. The throughput results derived from the simulation are then compared with the standard throughput of a Blockchain network operating without the DMI mechanism.

\subsubsection{\textbf{Simulation 1: Using the Default Setting in SIMBLOCK}}

In our initial analysis, simulation results based on SIMBLOCK default setting  to determine the initial forking rate against which the impact of incorporating DMI into the DTS strategy was evaluated, and to evaluate the throughput improvement. This methodology ensures that the baseline forking rate reflects the actual performance of the Blockchain network under default conditions. Moreover, since the simulation complies with the default setting in SIMBLOCK, simulation 1 can be compared with simulation 2-5 as a blank control group. In this simulation, we obtain a fork rate of 0.95\% and a throughput of 3.5 TPS.

\subsubsection{\textbf{Simulation 2:  DeFi Payment Without Applying DMI}}

\begin{figure}[H]
    \makeatletter
    \def\@captype{figure}
    \makeatother
    \centering
    \includegraphics[width=0.4 \textwidth]{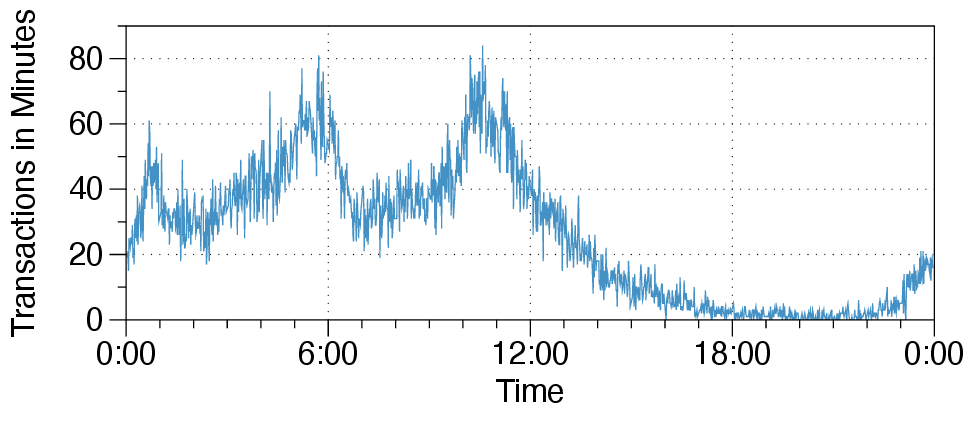}
    \caption{Payment transaction data obtained from the ICBC}
    \label{ICBC_raw}
\end{figure}

Next, we use the data set sourced from ICBC(Macau) (\reffig{ICBC_raw}), which contains 347071 payment data from July 2023 to August. This experiment's objective is to evaluate the efficacy of conducting payment transactions through the Blockchain without implementing DMI. \textcolor{black}{We establish the following settings for the Dynamic Transaction Selection (DTS) strategy: ``Scale'' is set to 6.8, ``Shape'' is set to 1.0, ``Transaction Incorporation Priority'' is defined as ``Time-based", ``Designated Space for Small Transaction'' is set to ``No'', and ``Maximum Space for One Transaction'' is capped at 80.}

\vspace{-0.3cm}

According to the data provided by ICBC (see \reffig{ICBC_raw}), the transaction volume reach up to 35760 per hour during the peak hour, while the transaction volume fall short of 530 during off-peak hours. Subsequently, the number transaction in transaction pool declined drastically as the addition of newly-made transaction can not catch up with the speed that transactions are packed into block.

The analysis of the transaction volume depicted in \reffig{ICBC_raw} unveils a discernible pattern over the course of the day. Commencing at midnight, the transaction volume commences at a relatively modest level of 530, encompassing fewer than 2100 individual transactions. As time progresses, the volume exhibits a gradual ascent from 530 at 00:00 to 25842 at 9:00, with minor oscillations. A significant inflection point is observed at 10:00, where the transaction volume experiences a surge to 33973. This increase is succeeded by a decline to 24271 by 15:00, followed by a resurgence to 35764 at 16:00. Finally, the day concludes with a transaction volume of 741.

\begin{figure}[htp]
    \makeatletter
    \def\@captype{figure}
    \makeatother
    \centering
    \includegraphics[width=0.4 \textwidth]{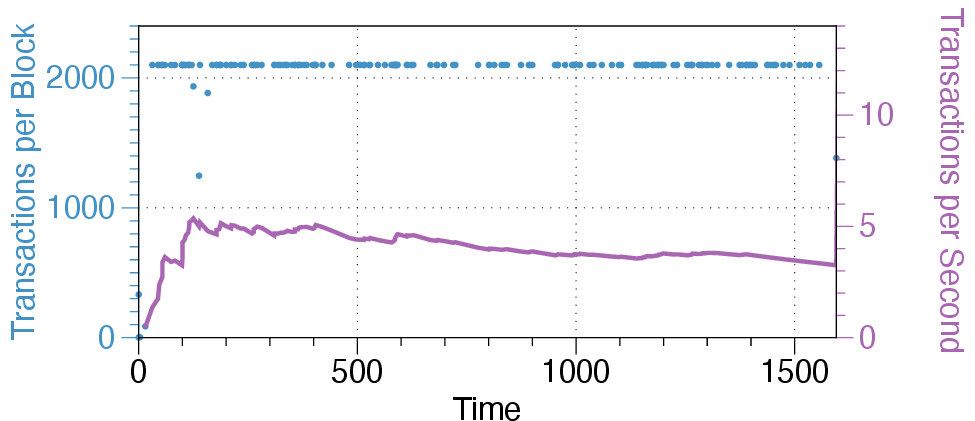}
    \caption{The block generation and throughput obtained from simulation 2}
    \label{withoutDMI}
\end{figure}

Building upon the insights derived from \reffig{ICBC_raw}, our simulation efforts are captured in \reffig{withoutDMI}. Evidently, the observed block generation pattern aligns with the initial hypotheses drawn from the raw data. The period from midnight to dawn sees a lower transaction volume, indicating a balance between transactions entering the transaction pool and those being included in blocks. Consequently, the initial three blocks manifest underfilled capacities. Concurrently, with the progressive increase in transaction volume, the transaction pool experiences rapid influx, subsequently leading to the majority of ensuing blocks being fully populated. The observed subtle fluctuations in this transition account for three additional blocks that remain partially filled. Notably, the mining of the last block at the 1595th minute, exceeding the span of 1440 minutes constituting a full day, results in the deferral of transactions from off-peak hours to the subsequent day. The calculated throughput in the scenario stands at 3.63 TPS, with a corresponding fork rate of 0.77\%.

\vspace{-0.3cm}

However, it becomes evident that the transaction processing capacity falters to adequately accommodate the surge in transaction volume during peak hours, leading to perceptible congestion within the transaction queue. This congestion contributes to the observed stable but low instant throughput, as shown in Fig.\ref{withoutDMI}. The dilemma between the demand for transaction processing and the system's inherent capacity during peak hours underscores the need for solutions to mitigate such congestion and enhance transaction processing efficiency.

\subsubsection{\textbf{Simulation 3: Applying DMI for DeFi Payment}}

\begin{figure}[H]
    \makeatletter
    \def\@captype{figure}
    \makeatother
    \centering
    \includegraphics[width=0.4 \textwidth]{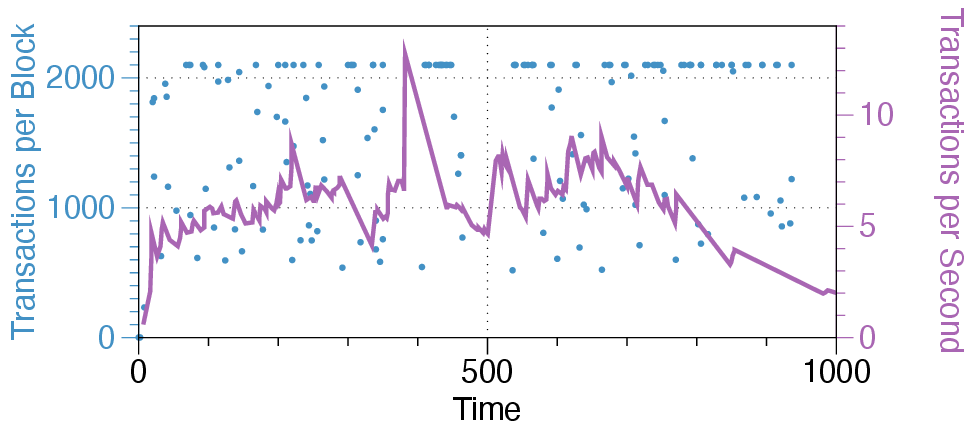}
    \caption{The block generation and throughput obtained from simulation 3}
    \label{withDMI}
\end{figure}

Upon implementing DMI within the DeFi system, a distinctive block generation pattern characterized by a notably denser distribution emerges, as evident in Fig.\ref{withDMI}. 

\vspace{-0.3cm}

Firstly, the integration of DMI exerts a pronounced influence on block mining dynamics. By strategically adjusting the target for blocks that are not fully filled, DMI engenders an environment wherein blocks are mined at a higher frequency. This proactive approach aligns with the aim of optimizing the system's throughput capabilities. Simultaneously, the reduced mining interval amplifies the pace at which transactions within the transaction pool are processed. This dual effect operates to reduce the likelihood of a block reaching full capacity, thereby ensuring a more stable and expeditious consumption of transactions. The mutually reinforcing interplay between these two facets culminates in a dynamic equilibrium that significantly augments throughput and mitigates the latency associated with pending transactions. Furthermore, the integration of DMI guarantees the timely inclusion of all transactions on the Blockchain within the confines of a single day, thus expediting the process of transaction verification and validation.

Consequently, the observed results within the scenario caused by DMI reveal a improvement in system performance. Specifically, the throughput reaches 4.46 TPS, reflecting a significant enhancement in transaction processing efficiency. The corresponding fork rate of 0.93\% reveal the stability and resilience of the Blockchain under DMI's influence, reinforcing the viability of this approach.

Of equal importance is the impact of DMI on eliminating congestion within the system. The evident resemblance in the instantaneous throughput pattern, as depicted in \reffig{withDMI}, to the payment data trends from ICBC highlights the efficacy of DMI in addressing congestion-related challenges. The convergence of peak activity around noon, coupled with the ensuing fluctuations, signifies a harmonious alignment between the Blockchain's transaction processing dynamics and the natural ebb and flow of transaction demand observed within real-world financial systems.

\subsubsection{\textbf{Comparison of Simulation 2 and 3}}
As detailed in TABLE \ref{BENCHMARK ACQUIRED FOR DeFi}, the application of DMI presents a 22.9\% surge in throughput, transitioning from 3.67 TPS to 4.46 TPS. This notable augmentation underscores DMI's favorable influence on DeFi's throughput efficiency. Beyond this advancement, the frequency of block generation exhibits improvement. Notably, in the absence of DMI, a mere 168 blocks are generated for the processing of all transactions, whereas with DMI in place, this count increases to 981 blocks. This observation underscores DMI's pivotal role in reducing the latency associated with transaction packaging, effectively mitigating pending times. This advancement transpires concurrently with the maintenance of a fork rate that remains below the 
baseline of 0.95

In summary, the empirical data presented provides a comprehensive validation of the positive impact of DMI on DeFi's performance metrics. The demonstrated increase in throughput and the faster block generation under DMI contribute significantly to the efficiency and effectiveness of the DeFi ecosystem. Importantly, the maintenance of a favorable fork rate showcases the robustness of the proposed DMI approach. These findings collectively suggest that the strategic incorporation of DMI has the potential to elevate DeFi networks, optimizing their performance, and fortifying their position as a transformative force within the financial landscape.

\vspace{-0.4cm}
\begin{table}[H]
  \centering
  \caption{\textcolor{black}{Simulation Results for Experiment 1}}
\small
  \label{BENCHMARK ACQUIRED FOR DeFi}
    \begin{tabular}{ c|c|c|c|c }
      \hline\hline
      \specialrule{0.00em}{3pt}{1pt} 
      Simulation & DTS & DMI & Throughput & Forking Rate\\
      \specialrule{0.00em}{3pt}{1pt} 
      \hline
      \specialrule{0.00em}{1pt}{1pt} 
      1 & \ding{56} & \ding{56} & 3.5 & 0.95\%\\
      \specialrule{0.00em}{1pt}{1pt} 
      \hline
      \specialrule{0.00em}{1pt}{1pt}  
      2 & \ding{51} & \ding{56} & 3.67 & 0.77\%\\
      \specialrule{0.00em}{1pt}{1pt} 
      \hline
      \specialrule{0.00em}{1pt}{1pt}  
      3 & \ding{51} & \ding{51} & 4.,46 & 0.93\%\\
      \specialrule{0.00em}{1pt}{1pt}   
      \hline
      \toprule
    \end{tabular}
\end{table}

\subsection{Experiment 2: Integrating DMI with DTS strategy for Cryptocurrency Transactions}

\begin{figure*}[htp]
    \centering
    \includegraphics[width=0.82 \textwidth]{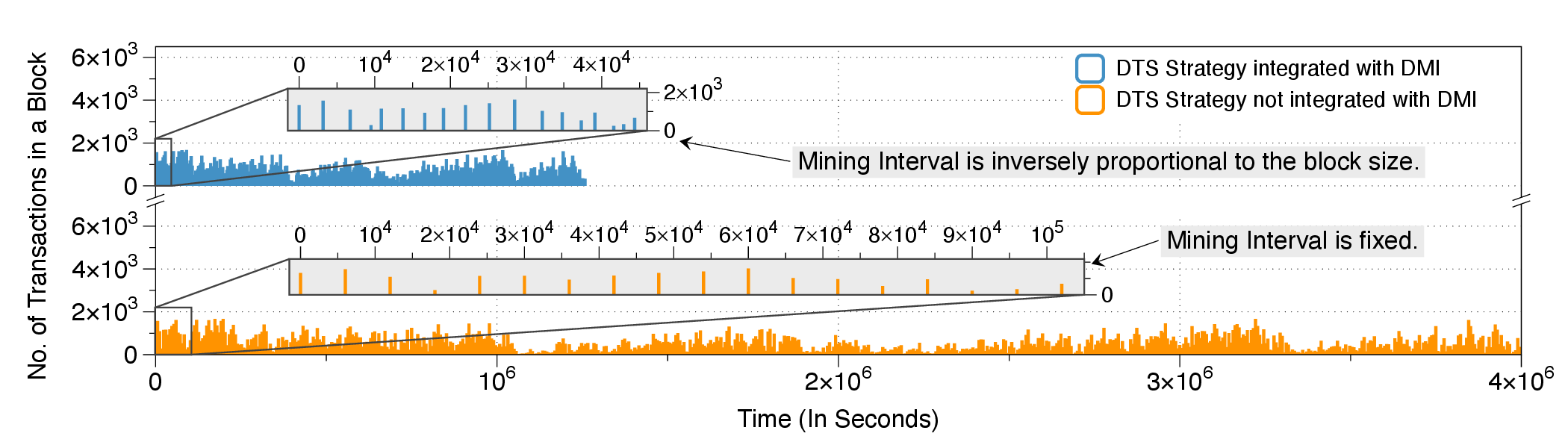}
    \caption{The histogram for simulations 4 and 5}
    \label{Target}
\end{figure*}

We use the historical market data spanning from January 2017 to December 2020, which contains 2,000,000 transactions, enriched with minute-to-minute updates encompassing OHLC (Open, High, Low, Close) values, transaction amounts denominated in BTC, and weighted Bitcoin price indices as detailed in \cite{BitcoinHistoricalData}. This experiment's objective is to evaluate the efficacy of DMI to address the inherent low throughput associated with the DTS strategy. The experimental framework encompassed the execution of two distinct simulations, each embodying a distinctive facet of the proposed methodology. 

\subsubsection{\textbf{Simulation 4: DTS strategy without using DMI}}
In this experiment, DMI is not applied when using DTS. The time used to packed all transaction was measured at 403,236.8 seconds. Significantly, this observed time closely aligns with the predicted duration of 402,000 seconds, thereby validating the accuracy of our time estimation methodology.

The experimental results are summarized in Table \ref{BENCHMARK ACQUIRED FOR DTS}. When we adopt DTS for the transactions, the resulting throughput was 0.94 TPS, which is notably lower when compared with the average throughput of 3.5 TPS achieved by the contemporary Bitcoin network. Evidently, this discrepancy underscores a notable reduction in throughput within the context of DTS, primarily attributed to its inherent characteristic of occupying a substantial amount of space within the system architecture.

Furthermore, fork rate was recorded at 0.2985\%. Notably, this fork rate reveals a low level of divergence in the Blockchain, indicating an avenue for optimizing the mining interval to potentially enhance throughput. The implications of this observation suggest that there exists an opportunity to strategically reduce the mining interval, thereby potentially achieving a heightened throughput. It is important to note that such an adjustment would inevitably introduce a certain degree of increase in the fork rate.

This experiment has provided valuable insights into the trade-off between throughput and fork rate. The demonstrated lower throughput of 0.94 TPS compared to the established benchmark of 3.5 TPS underlines the space-intensive nature of DTS, which substantiates its negative impact on throughput. However, the exploration of the simulation's fork rate of 0.2985\% illuminates a prospect for optimizing throughput by strategically shortening the mining interval, but with a measured acceptance of a rise in fork rate. This research highlights the nuanced interplay of these factors and paves the way for further investigation into enhancing Blockchain transaction efficiency.

\subsubsection{\textbf{Simulation 5: Integrating DMI with DTS Strategy}}
We integrate DMI with DTS strategy. The experimental results are summarized in Table \ref{BENCHMARK ACQUIRED FOR DTS}. The process of accommodating all transactions into blocks consumes a total of 106,442.6 seconds. Notably, this observed time frame closely aligns with the initially projected duration of 115,825.3 seconds, thereby validating the precision of our predictive modeling.

In this experiment, the resulting throughput was 3.57 TPS. This outcome holds substantial significance, as it positions the combined application of DTS and DMI in a commendable alignment with the current Bitcoin network. This similarity in throughput performance underscores the potential of the integrated DTS and DMI framework to rival the operational efficiency exhibited by the existing DTS strategy. Besides, the recorded fork rate was 0.895\%., which remains well below the baseline of 0.95\%. This observation underscores the robustness of the implemented methodology, further reinforcing the viability of the proposed DTS and DMI combination. The combination of DMI and DTS, as evidenced by the closely aligned transaction inclusion times and the commendable throughput rate of 3.57 TPS, underscores its potential to rival the existing operational benchmarks set by the Bitcoin network. 

\subsubsection{\textbf{Comparison of Simulation 4 and 5}}

As outlined in TABLE \ref{BENCHMARK ACQUIRED FOR DTS}, the application of DTS without the incorporation of DMI yields a throughput of 0.94 TPS. However, when DMI is subsequently introduced into the framework, the throughput reaches a level of 3.57 TPS. This surge in throughput signifies an amplification of 279\% in the system's capacity to efficiently handle transactions.

Note that the introduction of DMI does not alter the underlying mechanics of the DTS mechanism. As a result, the intrinsic characteristic of DTS, namely the maintenance of low volatility in block incentives, remains intact. This feature contributes to the sustained stability and predictability of block rewards, reinforcing the reliability of the DTS approach.

While this adjustment introduces a modest rise in the fork rate, it remains noteworthy that this increase is well-contained and situated below the established baseline of 0.95\%. This outcome highlights the robustness of the integrated DTS and DMI approach, as it maintains a commendable level of stability while concurrently driving substantial enhancements in throughput efficiency.

\vspace{-0.5cm}
\begin{table} [H]
    \centering
    \small
    \caption{\textcolor{black}{Simulation Results for Experiment 2}}
    \label{BENCHMARK ACQUIRED FOR DTS}
        \begin{tabular}{ c|c|c|c|c }
            \hline\hline
            \specialrule{0.00em}{3pt}{1pt} 
            Simulation & DTS & DMI & Throughput & Forking Rate\\
            \specialrule{0.00em}{3pt}{1pt} 
            \hline
            \specialrule{0.00em}{1pt}{1pt} 
            1 & \ding{56} & \ding{56} & 3.5 & 0.95\%\\
            \specialrule{0.00em}{1pt}{1pt} 
            \hline
            \specialrule{0.00em}{1pt}{1pt}  
            4 & \ding{51} & \ding{56} & 0.94 & 0.2985\%\\
            \specialrule{0.00em}{1pt}{1pt} 
            \hline
            \specialrule{0.00em}{1pt}{1pt}  
            5 & \ding{51} & \ding{51} & 3.37 & 0.895\%\\
            \specialrule{0.00em}{1pt}{1pt}   
            \hline
            \toprule
        \end{tabular}
\end{table}

\vspace{-0.3cm}
\textcolor{black}{The experimental results illustrated in \reffig{Target} reveal a significant decrease in the Blockchain network's throughput attributed to the existence of DTS. This emphasizes the importance of integrating DMI to overcome this limitation and enhance the network's overall performance. Upon implementing DMI in conjunction with DTS, notable improvements in throughput and a reduction in the mining interval are observed. These findings indicate that DMI effectively addresses the space-consuming drawback associated with DTS. Furthermore, it is worth noting that the volatility of the block incentive remains unchanged, indicating that the implementation of DMI preserves the advantageous aspects of DTS.}

\vspace{-0.3cm}
\subsection{Experiment 3: Comparison of DMI and Other Mechanism}
\begin{figure}[htp]
    \centering
    \includegraphics[width=0.4 \textwidth]{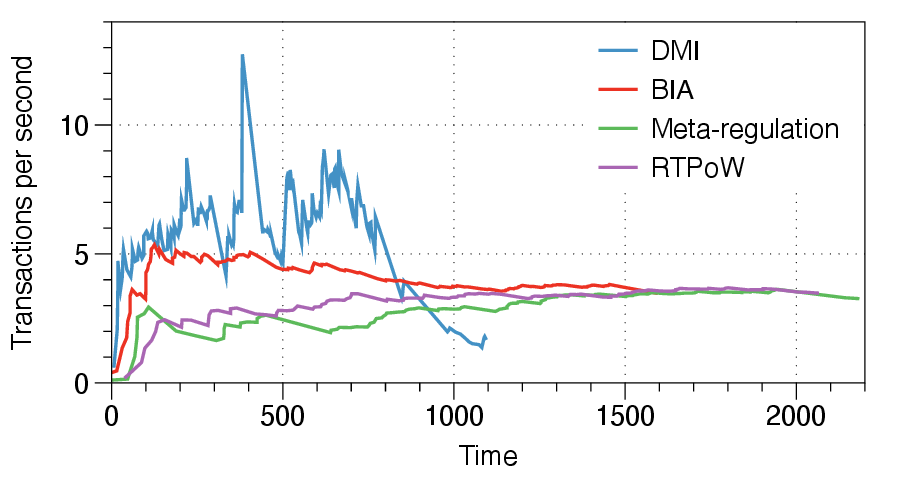}
    \caption{Applying DMI and other mechanisms for DeFi payment}
    \label{comparison}
\end{figure}

In this experiment, we implement BIA \cite{BlockIntervalAdjustment}, RTPoW \cite{RTPoW}, and Meta-regulation \cite{metaRegulation} for DeFi payments. The experimental results are shown in \reffig{comparison}.

 Note that our competitors also adhere to the same experimental configuration employed in DMI, utilizing the settings in TABLE \ref{SIMULATION SETTINGS BASED ON THE BLOCKCHAIN}. We observe that DMI out-performs BIA, RTPoW and Meta-regulation in terms of throughput, which show DMI's superiority in transaction processing for DeFi payment. It is because DMI is designed from the perspective of fork rate, block size and hash rate, while BIA, RTPoW and Meta-regulation are designed merely in concern of fork rate and hash rate. Therefore, under circumstances of stable hash rate and unstable block size, DMI is more efficient than the other mechanisms aforementioned. Additionally, we observe that the throughput of Blockchain when apply BIA is slightly higher than the throughput of Blockchain when applying RTPoW and Meta-regulation. It is because BIA also shortens the mining interval at the price of increased fork rate as DMI does.

\vspace{-0.4cm}
\begin{table}[htp]
  \centering
  \small
  \caption{\textcolor{black}{Comparison of DMI and other mechanisms}}
  \label{DMI comparison}
    \begin{tabular}{ c|p{4em}<{\centering}|p{4em}<{\centering}|p{7em}<{\centering} }
    \hline\hline
    \specialrule{0.00em}{1pt}{1pt} 
    Mechanism & Hash Rate & Fork Rate & Dynamically changing block\\
    \specialrule{0.00em}{1pt}{1pt} 
    \hline
    \specialrule{0.00em}{1pt}{1pt} 
    DMI & \ding{51} & \ding{51} & \ding{51}\\
    \specialrule{0.00em}{1pt}{1pt} 
    \hline
    \specialrule{0.00em}{1pt}{1pt} 
    RTPoW & \ding{51} & \ding{51} & \ding{56}\\
    \specialrule{0.00em}{1pt}{1pt} 
    \hline
    \specialrule{0.00em}{1pt}{1pt} 
    Meta-regulation & \ding{51} & \ding{51} & \ding{56}\\
    \specialrule{0.00em}{1pt}{1pt} 
    \hline
    \specialrule{0.00em}{1pt}{1pt} 
    BIA & \ding{51} & \ding{51} & \ding{56}\\
    \specialrule{0.00em}{1pt}{1pt} 
    \hline
    \toprule
\end{tabular}
\end{table}

In Table \ref{DMI comparison}, we compare the DMI and other mechanisms with respect to several key attributes. From Table \ref{DMI comparison}, we can observe that DMI mechanism serves as a comprehensive and versatile approach for Blockchain throughput optimization. It is strategically designed to bolster throughput, even at the expense of an elevated fork rate, to effectively handle fluctuations in transaction volume. When integrated with DTS strategy, the proposed DMI mechanism ensures stability in the fork rate, making it a fitting solution for both Decentralized Finance (DeFi) and cryptocurrency transactions. Unlike the Block Interval Adjustment (BIA) algorithm, DMI factors in block size variations to enhance throughput, also in periods of low fork rates. As a result, DMI presents a robust and effective strategy for navigating the complex dynamics of Blockchain while maximizing throughput efficiency.

\section{Conclusion}

In conclusion, the outcomes of this research establish the positive impact of DMI on overall throughput. \textcolor{black}{By applying DMI to DeFi payment, the throughput is increased from 3.67 TPS to 4.46 TPS}. Furthermore, this enhancement is complemented by a rise in the frequency of block generation. A comparison between scenarios with and without DMI reveals a clear distinction: while a mere 168 blocks sufficed for transaction processing without DMI, a significant 981 blocks were required with DMI, indicating a substantial reduction in transaction packaging latency attributable to DMI. Note that this improvement was achieved while comfortably maintaining a fork rate below the established baseline of 0.95\%.

\textcolor{black}{By implementing DTS to DMI, the throughput is increased from 0.94 TPS to 3.57 TPS.} This surge reflects a  enhancement in transaction handling capacity. Importantly, the introduction of DMI did not disrupt the underlying mechanism of DTS, thereby preserving the characteristic of low block incentive volatility obtained by DTS. Simultaneously, this improvement elevated the throughput performance to a level comparable to that of the current Bitcoin network. While a slight increase in the fork rate was observed due to the application of DMI, it remained well below the predefined baseline of 0.95\%.

In summary, these findings offer compelling evidence of DMI's potential in enhancing DeFi's throughput capabilities, expediting the frequency of block generation, and augmenting the efficacy of DTS, all while maintaining a low fork rate. These results underscore the potential significance of DMI as a vital component for the future enhancement and optimization of DeFi networks and the associated transaction systems. By leveraging the benefits of DMI, it is plausible that DeFi networks can achieve improved transaction efficiency and enhanced system performance, leading to a more robust and capable decentralized financial ecosystem.

\section*{Acknowledgments}
This research was funded by the University of Macau (file no. MYRG2022-00162-FST and MYRG2019-00136-FST).


    


\end{document}